\documentstyle[psfig,times,referee]{aa}      % A&A referee
\begin{document}

    \thesaurus{03       % A&A Section 3: Extragalactic astronomy
              (04.19.1;  % Surveys
               13.25.3   % X-rays: general
               13.25.2   % X-rays: galaxies
               13.25.5   % X-rays: stars
               11.01.2   % Galaxies: active
               11.03.1   % Galaxies: cluster: general 
             )}

\title
{
The Elliptical Galaxy formerly known as the Local Group: 
Merging the Globular Cluster Systems
} 

\titlerunning{Local Group Globular Cluster System}
\authorrunning{D. A. Forbes etal.}

\author{
Duncan A. Forbes \inst{1} \and Karen L. Masters \inst{1} \and 
Dante Minniti \inst{2} \and Pauline
Barmby \inst{3}
} 
\institute{School of Physics and Astronomy, 
University of Birmingham, Edgbaston, Birmingham B15 2TT, England  
\and 
Departamento de Astronom\'\i a y Astrof\'\i sica, P.~Universidad
Cat\'olica, Casilla 104, Santiago 22, Chile 
\and 
Harvard--Smithsonian Center for Astrophysics, 60 Garden St.,
Cambridge, MA 02138, USA
}

%\maketitle
\offprints{D. A. Forbes} 
\date{Received date; accepted date}

\maketitle

\begin{abstract}

Prompted by a new catalogue of M31 globular clusters, we have collected 
together individual metallicity values for globular clusters 
in the Local Group. Although we briefly describe the globular cluster 
systems of the individual Local Group galaxies, the main thrust of our
paper is to examine the collective properties. In this way we are
simulating the dissipationless merger of the Local Group, into presumably
an elliptical galaxy. 
Such a merger is dominated by the Milky Way and M31, which appear to be
fairly typical examples of globular cluster systems of spiral galaxies.  

The Local Group `Elliptical' has about 700 $\pm$ 125 globular clusters,
with a luminosity function resembling the `universal' one. The metallicity
distribution has peaks at [Fe/H] $\sim$ --1.55 and --0.64 with a 
metal--poor to metal--rich ratio of 2.5:1. The specific frequency of the 
Local Group Elliptical is initially about 1 but rises to about 3, when the
young stellar populations fade and the galaxy resembles an old elliptical. 
The metallicity distribution and stellar population corrected specific
frequency are similar to that of some known early type galaxies. 
Based on our results, we briefly speculate on the origin of globular cluster
systems in galaxies.


\end{abstract}

\section{Introduction}
\label{intro}
  The Local Group (LG) of galaxies currently consists of two large spirals (the
Milky Way and M31) and a host of smaller galaxies. Andromeda (M31) 
is approaching
the Milky Way with a velocity of about 60 km s$^{-1}$ 
and may ultimately collide and merge with our Galaxy (Dubinski et al. 1996). 
Such a collision of two near equal mass
spiral galaxies is expected to form an elliptical galaxy (Toomre \& Toomre
1972; Barnes \& Hernquist 1992). Globular clusters (GCs) are relatively robust
stellar systems and are expected to remain
intact during such a merger. Direct evidence for this comes from the GCs
associated with the Sgr dwarf galaxy which is currently being accreted
(e.g. Ibata et al. 1995; Minniti et al. 1996). Furthermore {\it Hubble Space
Telescope} studies of merging disk galaxies reveal evidence for the GC
systems of the progenitor galaxies (e.g. Forbes \& Hau 1999; Whitmore et al.
1999). 
Thus GCs should survive the merger of their parent galaxies, and will form
a new GC system around the newly formed elliptical galaxy. 

  The globular clusters of the Local Group are the best studied and offer a
unique opportunity to examine their collective properties (reviews of LG 
star clusters can be found in Brodie 1993 and Olszewski 1994). 
By examining them as a single GC system we are 
`simulating' the dissipationless (i.e. we assume that gas processes are
unimportant)  
merger of the two large spiral galaxies, their satellites and associated
galaxies. 
We note that the Milky Way and M31 are close in luminosity to 
L$^{\ast}$ galaxies, and so are highly relevant to a typical major 
merger in the local Universe.  
In reality, the merger of two spirals will involve gas which may lead to
the formation of new GCs (Ashman \& Zepf 1992). Here we do not speculate
about the number or properties of such new GCs but simply examine the
limited case of all currently existing Local Group 
GCs contributing to the new GC system. 

Thus our assumptions are: 1) that no GCs are destroyed in the merger (as
stated above, this is a reasonable assumption); 2)
that no GCs are created in the merger (in general, gaseous mergers will
create new clusters but here we are limiting ourselves to the simplier
case of no new GC formation); 3) the GCs without metallicity
determinations have a similar distribution to those measured (in practice
there will be a bias for missing GCs to be metal--rich as they are hidden
by inner bulges); 4) the missing GCs have a similar luminosity distribution
to the confirmed ones (our sample will be biased towards the more luminous
ones). 

In this paper, we collect together individual GC 
metallicities for all
known GC systems in the LG. 
Such a compilation is dominated by the GC
systems of the Milky Way and M31. 
We have used the June 1999 version of Harris (1996) for our Galaxy, and the 
recent catalogue of Barmby et al. (1999) for M31 but for the other 33
galaxies this required an extensive search of the literature. 
First we summarise the membership and properties of the Local Group
in the next Section. Section 3 briefly discusses the GCs in the 
individual LG galaxies. 

\section{The Local Group of Galaxies}
\label{The Local Group of Galaxies}

Reviews of the LG membership have been given by van den Bergh (1994a),
Grebel (1997), Mateo (1998) and Courteau \& van den Bergh (1999). 
Here we use the membership list of Courteau \& van den Bergh (1999). 

Within the Local Group, galaxies can be divided into three main subgroups. 
The first consists of the Milky Way and its satellites. 
This includes the Large and Small Magellanic Clouds, Fornax, and
Sagittarius as well as 9 other small dwarf galaxies. The second group 
consists of M31 and its satellites, the largest of which is M33, a spiral 
galaxy. The compact elliptical galaxy M32, the irregular galaxy IC 1613 and
numerous dwarf galaxies including NGC 147, NGC 185 and NGC 205 are also in the
M31 subgroup. Recently, two independent groups (Armandroff et al. 1998a,
1998b and Karachentsev \& Karachentseva 1999)  have found three new dwarf 
satellites of M31 named  And V, Pegasus II (And VI) and Cassiopeia (And
VII) which are included in our LG list. 
The third group is known as the Local Group Cloud (LGC), which is a
large cloud of mainly dwarf galaxies extending throughout the Local
Group. 
All of the galaxies in the Courteau \& van den Bergh
(1999) list are included in a subgroup, with 
2 galaxies having somewhat uncertain assignments (Mateo 1998).  

Galaxies that have at some point been associated with the Local Group but
are not on the Courteau \& van den Bergh (1999) list have not been included
here. Notable examples of these are the galaxies in the NGC 3109 (or
Antlia-Sextans) subgroup (NGC
3109, Antlia, Sextans A and Sextans B in Mateo 1998). Of these galaxies
NGC 3109 is the only one with evidence of a globular cluster system - Demers
et al. (1985) found ten globular cluster candidates. NGC 55 is another
example of a galaxy we have not included. This had been associated with
the LGC subgroup
(Mateo 1998) but seems more likely to actually belong in the Sculptor
(South Polar)
Group (Courteau \& van den Bergh 1999). NGC 55 has an estimated total
GC population of 25$\pm$15 (Liller \& Alcaino 1983) but only 3 with any
information (Da Costa \& Graham 1982; Beasley \& Sharples 1999). 
Other galaxies that have been
removed by Courteau \& van den Bergh (1999) include IC 5152 (which has 10
unconfirmed candidate GCs suggested in Zijlstra \& Minniti 1999), GR 8
which has no known GCs, and various other dwarf galaxies with no known GCs.

Simulations (e.g. Valtonen \& Wiren 1994) have suggested that the
IC 342/Maffei Group of galaxies (which consists of IC 342, Maffei 1 and 2,
Dwingloo 1 and 2, NGC 1569, NGC 1560, UGCA 105, UGCA 92, UGCA 86,
Cassiopeia 1 and MB 1; see Krismer et al. 1995) 
might once have been part of the LG but was
thrown out by interaction with M31. Courteau
\& van den Bergh (1999) do not include this group in the LG based
on their criteria for membership and we follow this assignment. It appears
that no study has been carried out of the
GC systems of galaxies in this group, although several galaxies might be
expected to have some based on their luminosity. We also mention
the Sab spiral 
M81 (NGC3031) in passing. Although not in the LG, at $\sim$ 3 Mpc 
it is sufficiently close for a detailed spectroscopic study. A
recent spectroscopic study using the Keck 10m telescope by
Schroder et al. (2000) finds that the GCs have similar spectra and 
line indices to Milky Way halo GCs. 

Our final list of LG galaxies is summarised in Table 1. We give galaxy
names, Hubble type, subgroup, distance, absolute V band 
luminosity and number of
GCs. Most properties come directly from Courteau \& van den Bergh (1999),
except the number of GCs which is the result of our investigation (see
Section 3).  
The Table is ordered by subgrouping within the Local Group.

\begin{table*}%[ht]
\caption[]{Local Group Galaxy Properties. 
Subgrouping is given by MW = Milky Way, M31 =
Andromeda, LGC = Local Group Cloud. Most quantities in this table are from
Courteau \& van den Bergh (1999). Distance is in kpc. 
See text for the number of globular clusters. 
A dash in the last column means that no
reliable mention of globular clusters in this 
galaxy could be found in the literature.
}
\begin{tabular}{llllrrr}
\noalign{\smallskip} \hline \noalign{\smallskip}
Name & Alternative Name  & Hubble Type & Subgroup & 
Dist. & M$_V$ & N$_{GC}$  \\
\noalign{\smallskip} \hline \noalign{\smallskip}
Milky Way &    & S(B)bc I-II & MW & 10 & --20.9 & 160$\pm$20\\
LMC &          & Irr III-IV  & MW & 50 & --18.5 & 19$\pm$16\\
SMC & NGC 292  & Irr IV/IV-V & MW & 60 & --17.1 & 8$\pm$7\\   
Sagittarius &  & dSph(t)     & MW & 30 & --13.8 & 4$\pm$1 \\
Fornax  &      & dSph        & MW & 140 & --13.1 & 5$\pm$0\\
Leo I & Regulus & dSph       & MW & 250 & --11.9 & --\\
Leo A & DDO 69 & dIrr V      & MW & 690 & --11.5 & --\\
Leo II & DDO 93 & dSph       & MW & 210 & --10.1 & --\\
Sculptor &     & dSph        & MW & 90  & --9.8  & --\\
Phoenix &      & dIrr/dSph   & MW/LGC & 400 & --9.8 & --\\
Sextans &      & dSph        & MW & 90  & --9.5  & --\\
Carina &       & dSph        & MW & 100 & --9.4  & --\\
Ursa Minor & DDO 199 & dSph  & MW & 60  & --8.9  & --\\
Draco & DDO 208 & dSph       & MW & 80  & --8.6  & --\\
M31 & NGC 224  & Sb I-II     & M31 & 760 & --21.2 & 400$\pm$55\\
M33 & NGC 598  & Sc II-III   & M31 & 790 & --18.9 & 70$\pm$15\\
M32 & NGC 221  & E2          & M31 & 760 & --16.5 & 0\\
NGC 205 & M110 & Sph         & M31 & 760 & --16.4 & 11$\pm$6\\
IC 10 & UGC 192 & Irr IV     & M31 & 660 & --16.3 & --\\
NGC 185 & UGC 396 & Sph      & M31 & 660 & --15.6 & 8$\pm$1\\
IC 1613 & DDO 8 & Irr V      & M31/LGC & 720 & --15.3 & 0\\
NGC 147 & DDO 3 & Sph        & M31 & 660 & --15.1 & 4$\pm$1\\
And I &        & dSph        & M31 & 810 & --11.8 & --\\
And II &       & dSph        & M31 & 700 & --11.8 & --\\
Pegasus II & And VI & dSph   & M31 & 830 & --10.6 & --\\ 
Pisces & LGS 3 & dIrr/dSph   & M31 & 810 & --10.4 & --\\
And III &      & dSph        & M31 & 760 & --10.2 & --\\
And V &        & dSph        & M31 & 810 & --10.2 & --\\
Cassiopeia & And VII & dSph  & M31 & 690 & --9.5  & --\\
NGC 6822 & DDO 209 & Irr IV-V & LGC & 500 & --16.0 & 1$\pm$1\\   
WLM & DDO 221  & Irr IV-V    & LGC & 930 & --14.4 & 1$\pm$1\\
Pegasus & DDO 216 & Irr V    & LGC & 760 & --12.3 & --\\
Aquarius & DDO 210 & Irr V   & LGC & 1020 & --11.3 & 1$\pm$1\\
SagDIG & UKS1927-177 & Irr V & LGC & 1400 & --10.7 & --\\
Tucana  &      & dSph        & LGC & 870 & --9.6  & --\\
\noalign{\smallskip} \hline
\end{tabular}
%\label{idents}
\end{table*}

\section{Globular Clusters of the Local Group}
\label{Globular Clusters of the Local Group}

\subsection{Luminosities}
\label{Luminosities}

Of the 35 LG galaxies, 13 of them are known to host GCs, giving a total 
number of GCs in the LG to be N$_{GC}$ = 692 $\pm$ 125. 
Individual V band magnitudes are available for a large number 
of these Local Group GCs. We have collected V band apparent magnitudes, 
irrespective of photometric error, for as many GCs as possible. The total 
number with V magnitudes available is 656, with about 2/3 of these coming from 
the M31 catalogue of Barmby et al. (1999). 
We note that the Barmby et al. list 
includes a number of GC candidates in addition to the confirmed
GCs. 

For some GCs it is not obvious which galaxy they should be associated
with. Although we try to assign the correct identity, it has little impact on
our final conclusions as we combine all GCs. 
We also note that none of the GCs, 
except those of the Milky Way, have been corrected for extinction. 
In the case of
the Milky Way the magnitudes can be reliably corrected using the 
reddening values quoted by Harris (1996; June 1999 version). 
All apparent magnitudes are converted 
into absolute magnitudes using the distances quoted in Table 1, or from 
Harris (1996) in the case of Milky Way GCs. 

We have chosen to exclude young clusters, i.e. with ages $<$ 3 Gyr or $(B-V)$
$<$ 0.6, as they may not evolve into {\it bona fide} globular clusters. 
From the point of view of comparison with elliptical galaxies, it is
reasonable to include only clusters older than $\sim$ 3 Gyr as elliptical
galaxies younger than this will tend to have extensive fine structure,
possibly tidal tails, and will probably not even be classified as an
elliptical galaxy. 

%In Fig. 1 we show the distribution of 
%absolute V band magnitudes for for all Local Group GC
%systems, with the number of individual luminosity 
%determinations available noted
%in each panel. The larger galaxies reveal a distribution peaked around 
%--6 $<$ M$_V$ $<$ --8, whereas the smaller galaxies generally have too few to 
%make conclusive statements. Interestingly the Fornax dwarf galaxy has a mean 
%M$_V$ that is very similar to M31 and the Milky Way, even though it only has 
%5 GCs. 

In Fig. 1 we show the combined V band luminosity for 656 Local Group GCs out 
of a total of 692 $\pm$ 125. The histogram is dominated by M31 GCs, which
have not been extinction corrected, and therefore the peak of the
distribution is somewhat fainter than for the Milky Way alone (e.g. 
Della Valle et al. 1998). 
Nevertheless the distribution resembles a Gaussian with a
dispersion of $\sim$1 mag, and is similar to those seen in 
many elliptical galaxies (e.g. Whitmore 1997 for a review).

\subsection{Metallicities}
\label{Metallicities}

The main aim of this paper is to examine the overall metallicity
distribution of LG globular clusters. Unfortunately
there are fewer individual 
GC metallicities available in the literature than magnitudes. 
We have found a total of 
388, which represents over half of the suspected total number of 
692 $\pm$ 125 GCs.
The bulk of these 
metallicity determinations come from spectroscopy or colour--magnitude 
diagrams. However for a number of GCs we only have colour information
available; typically $(B-V)$. In these cases 
we have used the Galactic [Fe/H]--colour relation as derived by Barmby
et al. (1999). They used only GCs with low extinction, i.e. $E_{B-V}<0.5$, 
from  the latest Harris (1996) compilation. 
Colours are de-reddened using the Harris values for $E_{B-V}$ and the 
extinction curve of Cardelli, Clayton \& Mathis (1989). For colours we 
have assigned a characteristic 
error in [Fe/H] of 0.5 dex. In the half dozen cases where 
the colours suggest (an implausible) 
[Fe/H] $>$ 1 we have set the metallicity to [Fe/H] = 1.0.

The Barmby et al. Galactic relation assumes a linear relation, although
there is some question whether this is valid at the high metallicities (see
Barmby et al. 1999 for discussion). 
We also note that Kissler--Patig et al. (1998) have derived a new
[Fe/H] vs $(V-I)$ relationship based on spectroscopic metallicities for 
GCs in the giant elliptical NGC 1399. This has the advantage of extending
the metallicity range to GCs more metal-rich than typically found in the
Milky Way. They find that the (linear) slope of the 
relation (3.27 $\pm$ 0.32) is
almost twice as flat as conventional (Galactic) fits. Barmby et al. 
derive a $(V-I)_o$
slope of 4.22 $\pm$ 0.39. So for red GCs, metallicities derived from
relations based on Galactic fits tend to be overestimated. Although this is
a serious effect for the GC systems of 
elliptical galaxies (which tend to have fairly red median colours), it is
much less important for late type galaxies, as found in the Local Group. 
Thus here we use the Galactic linear fits, which should be reasonable for
our purposes. 

Our compilation is dominated by the Milky Way and M31. For these we use the
lists of Harris (1996; June 1999 version) and Barmby
et al. (1999) respectively. For M33 we list individual GC metallicities
in Table 2 and in Table 3  we list GC metallicities for the remaining LG
galaxies.

\begin{table*}%[ht]
\caption[]{M33 Globular Cluster Metallicities. 
S = spectroscopy, CMD = colour-magnitude diagram, C
= colour. 1 = Christian \& Schommer 1988, 2 = Brodie \& Huchra 1991, 3 =
Sarajedini et al. 1998, a = assumed metallicity. 
}
\begin{tabular}{l c c c c}
\noalign{\smallskip} \hline \noalign{\smallskip}
Name &
$[$Fe/H$]$ &
$\sigma$($[$Fe/H$]$) &
Source &
Ref. \\
\noalign{\smallskip} \hline \noalign{\smallskip}
U49 & -1.43 & 0.30 & S/CMD & 1,2,3\\
R13 & -0.75 & 0.5 & C     & 1\\
R12 & -1.19 & 0.27 & S/CMD & 1,3\\
R15 & -1.79 & 0.5 & C     & 1\\
R14 & -1.37 & 0.41 & S/CMD & 1,3\\
M9  & -1.67 & 0.29 & S/CMD & 1,3\\
U77 & -1.59 & 0.58 & S/CMD & 2,3\\
H38 & -1.14 & 0.22 & S/CMD & 1,3\\
H21 & -0.40 & 0.40 & S     & 1\\
C20 & -1.57 & 0.50 & S/CMD & 1,2,3\\
C36 & -1.00 & 0.30 & S     & 1\\
C38 & -0.77 & 0.24 & S/CMD & 1,3\\
C18 & -0.46 & 0.56 & S/CMD & 1,2\\
H10 & -1.40 & 0.66 & S/CMD & 2,3\\
C3  & -2.38 & 0.56 & S     & 2\\
C32 & -1.77 & 1.12 & S     & 2\\
U137 & -0.85 & 0.29 & S/CMD & 2,3\\
C21 & -1.85 & 0.5 & C & 1\\
U23 & -0.91 & 0.5 & C & 1\\
C9  & -2.12 & 0.51 & S     & 2\\
S24 & +0.52 & 0.5 & C & 1\\
S72 & -1.30 & 0.5 & C & 1\\
S247 & +1.0$^a$ & 0.5 & C & 1\\
U67 & -1.74 & 0.5 & C & 1\\
S161 & -1.90 & 0.5 & C & 1\\
S160 & +0.85 & 0.5 & C & 1\\
U7  & -1.52 & 0.5 & C & 1\\
\noalign{\smallskip} \hline
\end{tabular}
%\label{idents}
\end{table*}

\begin{table*}%[ht]
\caption[]{Globular Cluster Metallicities. 
S = spectroscopy, CMD = colour-magnitude diagram, C
= colour. 1 = Suntzeff et al. 1992, 2 = Olsen et al. 1998, 3 = Dutra et al.
1999, 4 = Sarajedini 1998, 5 = Olszewski et al. 1991, 6 = Da Costa \&
Hatzidimitriou 1998, 7 = Mighell et al. 1998, 8 = Harris 1996, 9 = Buonanno
et al. 1998, 10 = Dubath et al. 1992, 11 = Barmby et al. 1999, 12 = Da Costa
\& Mould 1988, 13 = Cohen \& Blakeslee
1998, 14 = Hodge et al. 1999, 15 = Greggio
et al. 1993, a = assumed error, b = assumed metallicity.
}
\begin{tabular}{l c c c c}
\noalign{\smallskip} \hline \noalign{\smallskip}
Name &
$[$Fe/H$]$ &
$\sigma$($[$Fe/H$]$) &
Source &
Ref. \\
\noalign{\smallskip} \hline \noalign{\smallskip}
 & & LMC & & \\
\hline
Hodge II & -2.06 & 0.20 & S & 1\\
NGC 1466 & -1.85 & 0.20 & S & 1\\
NGC 1754 & -1.46 & 0.18 & S/CMD & 1,2\\
NGC 1786 & -1.87 & 0.20 & S & 1\\
NGC 1835 & -1.68 & 0.18 & S/CMD & 1,2\\
NGC 1898 & -1.25 & 0.18 & S/CMD & 1,2\\
NGC 1916 & -2.08 & 0.20 & S & 1\\
NGC 2005 & -1.57 & 0.18 & S/CMD & 1,2\\
NGC 2019 & -1.44 & 0.18 & S/CMD & 1,2\\
NGC 2210 & -1.97 & 0.20 & S & 1\\ 
NGC 2257 & -1.80 & 0.10 & CMD & 1\\  
NGC 1841 & -2.11 & 0.10 & S & 1\\
Reticulum &  -1.71 & 0.10 & S & 1\\ 
NGC 1928 &    -1.2 & 0.3$^a$ & S & 3\\
NGC 1939 &    -2.0 & 0.3$^a$ & S & 3\\
ESO 121-SC03 & -0.96 & 0.13 & S/CMD & 4,5\\
NGC 2121 &    -1.04 & 0.13 & CMD & 4\\ 
SL 663  &     -1.05 & 0.16 & CMD & 4\\
\medskip
NGC 2155 &    -1.08 & 0.12 & CMD & 4\\

 & & SMC & & \\
\hline
NGC 221 & -1.48 & 0.11 & CMD/S & 6,7\\
Lindsay1 &   -1.21 & 0.10 & CMD/S & 6,7\\
Kron3  &     -1.10 & 0.11 & CMD/S & 6,7\\
Lindsay113 & -1.26 & 0.13 & CMD/S & 6,7\\
NGC 339 &     -1.38 & 0.12 & CMD/S & 6,7\\
NGC 416 &     -1.44 & 0.12 & CMD & 7\\
NGC 361 &     -1.45 & 0.11 & CMD & 7\\
\medskip
Lindsay11 &  -0.81 & 0.14 & CMD/S & 6,7\\ 

 & & Sagittarius & & \\
\hline
NGC 6715/M54 & -1.59 & 0.11 & S & 8\\ 
Ter7 &       -0.58 & 0.09 & S & 8\\
Arp2 &       -1.76 & 0.10 & S & 8\\
\medskip
Ter8 &       -2.00 & 0.12 & S & 8\\
\noalign{\smallskip} \hline
\end{tabular}
\end{table*}

\begin{table*}%[ht]
%\caption[]{Globular Cluster Metallicities. 
\begin{tabular}{l c c c c}
\noalign{\smallskip} \hline \noalign{\smallskip}
 & & Fornax & & \\
\hline
Fornax-1 & -2.20 & 0.20 & CMD & 9\\
Fornax-2 & -1.78 & 0.20 & CMD & 9\\
Fornax-3 & -1.94 & 0.18 & CMD/S & 9,10\\
Fornax-4 & -1.55 & 0.18 & CMD/S & 9,10\\
\medskip
Fornax-5 & -2.00 & 0.18 & CMD/S & 9,10\\

 & & NGC 205 & & \\
\hline
Hubble I/009-061 & -1.52 & 0.21 & C/S & 11,12\\
Hubble II/011-063 & -1.51 & 0.26 & C/S & 11,12\\
Hubble IV/328-054 & -1.58 & 0.22 & C/S & 11,12\\     
Hubble VI/331-057 & -1.30 & 0.49 & C/S & 11,12\\
Hubble VII/330-056 & -1.40 & 0.10 & S & 12\\
\medskip
Hubble VIII/317-041 & -1.93 & 0.28 & C/S & 11,12\\

 & & NGC 147 & & \\
\hline
NGC 147-1 & -1.90 & 0.15 & S & 12\\  
\medskip
NGC 147-3 & -2.50 & 0.25 & S & 12\\

 & & NGC 185 & & \\
\hline
NGC 185-1 & -1.40 & 0.10 & S & 12\\
NGC 185-2 & -1.20 & 0.25 & S & 12\\
NGC 185-3 & -1.70 & 0.15 & S & 12\\
NGC 185-4 & -2.50 & 0.25 & S & 12\\
\medskip
NGC 185-5 & -1.80 & 0.15 & S & 12\\

 & & NGC 6822 & & \\
\hline
\medskip
NGC 6822-VII & -1.95 & 0.15 & S & 13\\

 & & WLM & & \\
\hline
\medskip
WLM-1 & -1.52 & 0.08 & CMD & 14\\

 & & Aquarius & & \\
\hline
\medskip
Aquarius-1 & +1.0$^b$ & 0.5 & C & 15\\
\noalign{\smallskip} \hline
\end{tabular}
%\label{idents}
\end{table*}

Figure 2 shows the metallicity distributions for all Local Group GC
systems, with the number of individual [Fe/H] determinations available noted
in each panel. It suggests multiple peaks in the distribution of 
several galaxies, with only the three LG spirals having a substantial
population of relatively metal--rich GCs. 
In Table 4 we summarise the LG galaxies with GCs, giving their mean
metallicity and the total specific
frequency. 

\begin{table*}%[ht]
\caption[]{Local Group Galaxy Properties. 
$[$Fe/H$]$ clusters gives the mean metallicity of the
globular cluster system. 
The specific frequency S$_N = N_{GC} \times 10^{0.4(M_V +
15)}$, error based on uncertainty in N$_{GC}$ only. 
%a = Ashman \& Bird 1993, b = Barmby et al. 1999.
}
\begin{tabular}{l c c c}
\noalign{\smallskip} \hline \noalign{\smallskip} 
Name &
$[$Fe/H$]$ &
S$_N$  \\
 &
clusters &
\\
\noalign{\smallskip} \hline \noalign{\smallskip} 
Milky Way   & -1.59, -0.55   &  0.7$\pm$0.1 \\
LMC         & -1.58          &  0.8$\pm$0.6 \\
SMC         & -1.29          &  1.2$\pm$1.0 \\ 
Sagittarius & -1.38          & 12.0$\pm$3.0 \\
Fornax      & -1.89          & 28.8$\pm$0.0 \\ 
M31         & -1.40, -0.58    &  1.3$\pm$0.2 \\
M33         & -1.32          &  1.9$\pm$0.4 \\
NGC 205     & -1.48          &  3.0$\pm$1.7 \\ 
NGC 185     & -1.61          &  4.6$\pm$0.6 \\
NGC 147     & -2.06          &  3.6$\pm$0.9 \\
NGC 6822    & -1.95          &  0.4$\pm$0.4 \\
WLM         & -1.52          &  1.7$\pm$1.7 \\ 
Aquarius    & +1.0           &  30.2$\pm$30.2 \\
\noalign{\smallskip} \hline
\end{tabular}
%\label{idents}
\end{table*}

\subsection{The Milky Way}
\label{The Milky Way}

   The Milky Way galaxy has 147 globular clusters listed in the compilation
of Harris (1996). However the discovery
of the dSph galaxy Sagittarius
(Ibata et al. 1995) has raised questions as to whether 4 of these clusters 
(NGC 6715/M54, Terzan 7, Arp 2 and Terzan 8) should actually be associated with
Sagittarius. There has been some discussion in the literature  about the
inclusion of Terzan 7 in this list (e.g. Minniti et al. 1996), but van den
Bergh (1998) argues that it probably should be associated with the 
Sgr dwarf so we will remove it and the other three 
from the Milky Way GC system.

The presence of  
younger, counter--rotating halo GCs (Zinn 1993; Majewski
1994) indicates that the Milky Way has accreted other 
small galaxies and their GC systems in the past.
In particular, 
Unavane et al. (1996) have suggested that $\le$ 6 Sgr or Fornax like
dwarfs have been accreted over the last 10 Gyrs. This implies
that less than 30
GCs are `foreign' to the Milky Way system, and 
that past accretions of GCs are fairly rare. 

Individual GC metallicities come from the latest (i.e. June 1999) 
electronic version of Harris (1996) which uses CMD diagrams and
spectroscopy (see Harris 1996 for details). All but two GCs (BH 176 and
Djorg 1) have listed metallicities. For these GCs we have estimates of the 
metallicities from Ortolani et al. (1995a) for BH 176 and Ortolani et al.
(1995b) for Djorg 1. Both have [Fe/H] $\sim$ --0.4. This gives a total of
143 MW GCs with metallicities. Some metal--rich GCs are no doubt 
hidden from the Sun's position (Minniti 1995), 
and we adopt a total system population of 
160$\pm$20 globular clusters from van den Bergh (1999). 

A KMM analysis  
of the metallicity distribution shown in Fig. 2 
indicates a bimodal metallicity distribution 
with peaks at [Fe/H] = --1.59 and --0.55, which is consistent with the 
findings of Ashman \& Bird (1993) who found 
[Fe/H] $\sim$ --1.6 and --0.6.
We find that the ratio of 
metal--poor to metal--rich GCs is 2:1.

\subsection{M31}
\label{M31}

  Andromeda is the largest galaxy in the Local Group and has 14$\pm$1 
companions (Courteau \& van den Bergh 1999). Recently a new catalogue of
the globular cluster system in M31 has been
compiled (Barmby et al. 1999). It includes 437 clusters and cluster
candidates, of which 162 have metallicities from spectroscopy and 90 more have
metallicities calculated from their colours. 
Twelve of the clusters are flagged as possibly being members of NGC 205, 
six of 
which have metallicites given. 
These 12 will be taken to be associated with NGC 205 as described below.
We have removed the 59 clusters with $(B-V)$ $<$ 0.55, as these may not be
{\it bona fide} GCs. 
As done in Barmby et al., we only use 
those GCs with metallicity errors $<$ 0.5 dex to give a total of 165,
121 of which have metallicities from spectroscopy. 
In this case, the colours of the M31 GCs were converted into
[Fe/H] estimates using the transformation derived by Barmby et al. (1999)
based on an extinction model and direct measurements of the M31 system.  
The total number of GCs in the M31 GC system is taken to be 
400 $\pm$ 55 from van den Bergh (1999). 

A KMM analysis 
of the metallicity distribution shown in Fig. 2 
indicates a bimodal metallicity distribution 
with peaks at [Fe/H] = --1.40 and --0.58. This is similar to Ashman \& Bird 
(1993) who found [Fe/H] $\sim$ --1.50 and --0.6, based on the earlier data of 
Huchra, Brodie \& Kent (1991) (which forms a part of the Barmby et al. 1999 
data set). The ratio of metal--poor to metal--rich GCs is 3:1.

\subsection{NGC 205}
\label{NGC 205}

  There has been much debate in the literature about exactly how
many GCs are associated with NGC 205 and which belong with 
M31. NGC 205 is close in position to M31 and has a radial velocity within the
range given for the internal dispersion of the M31 
GC system (Reed et al. 1992). 
This makes it difficult to determine which galaxy the GCs 
should  be associated with, and is further complicated by the fact that
NGC 205 appears to be interacting with M31 (Zwicky 1959). 
  Barmby et al. (1999) flag 12 of the GCs in their catalogue 
as possibly being associated with NGC 205. They give metallicities for six
of these. Da Costa \& Mould (1988) give the metallicity for five of these
plus an additional one (Hubble VII/330-056), giving a total of seven
clusters with metallicities
available. 
We have decided to adopt the Barmby et al.
(1999) list of 12, but have removed Hubble V/324-051 as it is relatively 
blue and has a 
spectrum with strong Balmer lines. Da Costa \& Mould (1988)
also note that it may be a young cluster.
For the total number of GCs associated with NGC 205 we adopt 11 $\pm$
6.

\subsection{M33}
\label{M33}

Three subpopulations of clusters may be present in
M33. The blue clusters follow the disk
rotation, the red ones show no evidence for rotation and have a large
line-of-sight velocity dispersion, while intermediate coloured ones have
intermediate kinematic properties. The blue subpopulation are thought to be
young and the red ones forming an old spherical halo (Schommer et al.
1991). Sarajedini et al. (1998) summarise the evidence that the halo 
clusters in M33 formed over
a long time span. 
Christian \& Schommer (1988) identify 27 likely candidates for true
globular clusters in M33 on the basis that they have $B-V > 0.6$. 
Chandar, Bianchi \& Ford (1999) list 60 M33 stellar clusters, 49 of 
which they refer 
to as `populous', meaning that they had a shape similar to that expected for
globular clusters. However, we have not included these clusters in the
metallicity distribution as it is not clear which objects are 
genuine GCs. The total number of GCs has been estimated
by R. Chandar as part of her thesis on M33 to be 70 $\pm$ 15 (Chandar
1999). 

\subsection{The Large Magellanic Cloud}
\label{The Large Magellanic Cloud}

  The Large Magellanic Cloud (LMC) GC system contains a population of 
genuine old ($\sim$ 10 Gyr) globular clusters and a class of young ($\le$ 3
Gyr) clusters (van den Bergh 1994b). 
The only exceptions to this rule appear to be ESO 121-SC03 which
has an age of around 9 Gyr (Dutra et al. 1999) and three GCs 
recently found to have ages of around 4 Gyr (Sarajedini 1998).
  
Suntzeff et al. (1992) reviewed thirteen {\it bona fide} old clusters (meaning
clusters of similar age to Galactic globular clusters) which can be found
in the Large Magellanic Cloud, giving spectroscopic metallicities for all
but one, for which we have derived a metallicity from its 
$B-V$ colour.  Geisler et al. (1997) 
conducted a search to find more old clusters in the LMC and concluded that 
``there are few, if any, genuine old clusters in the LMC left to be
found.'' However Dutra et al. (1999) present spectroscopic evidence that NGC
1928 and NGC 1939 are also old globular clusters in the LMC taking the
census up to 15. Olsen et al. (1998) 
produced CMDs for five of the old globular clusters (NGC
1754, NGC 1835, NGC 1898, NGC 2005 and NGC 2019), and claim
that the metallicites they derive from the CMDs are more accurate than the
earlier spectroscopic determinations.  

It is not clear whether the young (i.e. $\le$ 3 Gyr) LMC clusters will
eventually evolve into {\it bona fide} GCs. Here we have decided not to
include them in our compilation. Thus we adopt the 15 old GCs and the 4
intermediate aged ones to give a total GC system of 19, which could 
perhaps be as high as 35. 

\subsection{The Small Magellanic Cloud}
\label{The Small Magellanic Cloud}

  The Small Magellanic Cloud (SMC), like the LMC has both old and
intermediate aged globular clusters. However unlike the LMC it has several
(at least 7, Mighell et al. 1998)
GCs between the ages of 3 and 10 Gyr. There is only one old globular
cluster (NGC 121) with an age of $\sim$ 11 Gyr. 
We include this GC plus the 7 intermediate
aged GCs. Spectroscopic metallicities exist for six of them 
(Da Costa \& Hatzidimitriou  1998), for the other two we estimate [Fe/H]
from their colours.

\subsection{Sagittarius}
\label{Sagittarius}

The Sagittarius dwarf has four globular clusters (Ibata et al. 1995) which were
previously thought to be members of the Galactic GC 
system (as described in section 3.3). We have metallicities for all of
these. One of them (NGC 6715/M54) is the second brightest
globular cluster in the Milky Way and is located close to the centre of
Sagittarius. This has led to the idea that it might actually be the nucleus of
Sagittarius, as discussed in Mateo (1998); 
however we will include it as a
globular cluster.

\subsection{Fornax, NGC 147 and NGC 185}
\label{Fornax, NGC 147 and NGC 185}

The Fornax dSph galaxy has five globular clusters, 
all of which have metallicities available from their CMDs (Buonanno et al.
1998, 1999) and three of which also have spectroscopic metallicities
(Dubath et al. 1992). The metallicities listed in Table 3 are the weighted
average.   
  The dwarf elliptical NGC 147 has four globular clusters (Minniti et al.
1996), two of which have metallicities available from spectroscopy (Da
Costa \& Mould 1988).
  The dwarf elliptical NGC 185 has eight globular clusters (Minniti et al.
1996), five of which have spectroscopic metallicities available (Da Costa \&
Mould 1988).

\subsection{WLM, NGC 6822 and Aquarius}
\label{WLM, NGC 6822 and Aquarius}

  The dwarf irregulars WLM, NGC 6822 and Aquarius have one globular cluster
each (Harris 1991 for WLM and NGC 6822, Greggio et al. 1993 for Aquarius).
Recently, Hodge et al. (1999) have determined the metallicity of the WLM 
GC to be [Fe/H] = --1.52 $\pm$ 0.08 from fitting isochrones to its CMD. 
The metallicity of the 
NGC 6822 GC from spectroscopy is [Fe/H] = --1.95 $\pm$ 0.15
(Cohen \& Blakeslee 1998). The colour of the GC in Aquarius is 
$B-V$ = 1.15 (Greggio et al. 1993) which corresponds to a
metallicity of +1.07 $\pm$ 0.5 (we have set this to a value of +1.0 in Table
3).

\subsection{Galaxies with no Globular Cluster System}
\label{Galaxies with no Globular Cluster System}

  No globular clusters have been found in the dwarf irregular galaxy IC 1613
(M$_V$ = --14.7) or the elliptical galaxy M32 (Harris 1991). In the case of
the compact elliptical M32, its outer stars, and presumably any GCs, may have
been stripped away by interaction with M31. Given its
proximity to M31, it is very difficult to form a definitive
conclusion on the GC population of M32. 

 No information has been found in the literature on the GC system of the
remaining galaxy in the Courteau \& van den Bergh (1999) list, i.e.   
the dwarf
irregular IC 10 with M$_V$ = --16.3. It is more luminous than  
Fornax and Sagittarius but has no GCs found to date. 
Such a search may prove fruitful. 

Except for Aquarius (at M$_V$ = --11.3), no GCs have been found
in any LG galaxy fainter than 
Fornax and Sagittarius (M$_V$ = --13.1 and 
--13.8 respectively). 
This applies
to 19 low luminosity 
galaxies in the Local Group (see Table 1).

%\section{NGC 4565: Another Spiral}
%\label{NGC 4565: Another Spiral}

%Perhaps one of the best observed GC systems in a spiral galaxy, 
%outside of the LG, is NGC 4565 (Kissler--Patig et al. 1999). This 
%edge-on Sb galaxy (M$_V$ = --21.4) was recently imaged 
%with the HST in the B and I bands. From the 40 
%detected GCs, Kissler--Patig et al. estimate a 
%total GC population of 204 $\pm$ 38, giving S$_N$ = 0.56 $\pm$ 0.15. 
%These properties make NGC 4565 and its GC system comparable to M31, and to 
%some extent the Milky Way (see Table 1). Its specific frequency is also
%comparable to other late type spirals (see Table 5).  
%It is also one of the few spiral 
%galaxies outside of the LG to have individual GC colours available. 
%In Fig. 3 we show its metallicity distribution, based on our Galactic
%$(B-I)$ colour transformation. There is a suggestion of two peaks, around 
%[Fe/H] $\sim$ --1.3 and --0.5, although it is not statistically significant 
%according to an KMM analysis. The distribution is similar to that of 
%M31 and the MW as shown in Fig. 2. 

\begin{table*}%[ht]
\caption[]{Specific Frequencies of Spiral Galaxies. 
1 = Kissler-Patig et al. (1999), 2 = Ashman \& Zepf (1998).
}
\begin{tabular}{l l l l r r r}
\noalign{\smallskip} \hline \noalign{\smallskip} 
Name &
Hubble Type &
M$_V$ &
S$_{N}$ &
Ref.\\
\noalign{\smallskip} \hline \noalign{\smallskip} 
LG       &  E          & -22.0 & 1.1$\pm$0.2 & This work\\ 
LG (Age corrected) &  E & -20.9 & 3.0$\pm$0.5 & This work\\
Milky Way & S(B)bc      & -20.9 & 0.7$\pm$0.1 & This work\\
M31 &       Sb          & -21.2 & 1.3$\pm$0.2 & This work\\
\medskip
M33 &       Sc          & -18.9 & 1.9$\pm$0.4 & This work\\
NGC 4565 &  Sb          & -21.4 & 0.56$\pm$0.15 & 1 \\ 
NGC 5907 &  Sc          & -21.2 & 0.56$\pm$0.17 & 1 \\
NGC 253  &  Sc          & -20.2 & 0.2$\pm$0.1 & 2 \\
NGC 2683 &  Sb          & -20.8 & 1.7$\pm$0.5 & 2 \\
NGC 3031 (M81) &  Sab         & -21.1 & 0.7$\pm$0.1 & 2 \\
NGC 4216 &  Sb          & -21.8 & 1.2$\pm$0.6 & 2 \\
NGC 4569 &  Sab         & -21.7 & 1.9$\pm$0.6 & 2 \\
NGC 4594 &  Sa          & -22.2 & 2$\pm$1 & 2 \\
NGC 5170 &  Sb          & -21.6 & 0.9$\pm$0.6 & 2 \\
NGC 7814 &  Sab         & -20.4 & 3.5$\pm$1.1 & 2 \\
\noalign{\smallskip} \hline
\end{tabular}
\label{idents}
\end{table*}

\section{Properties of the Local Group `Elliptical'}
\label{Properties of the Local Group `Elliptical'}

In Fig. 3 we show 
the combined metallicity distribution for 387 Local Group GCs 
with available metallicities. A KMM analysis, excluding the 6 outliers, 
indicates two 
peaks at [Fe/H] =  --1.55 and --0.64, with the ratio of metal--poor to 
metal--rich GCs being 2.5:1. This is likely to slightly 
overestate the ratio; the  
missing GCs are likely to be found 
in the central bulge 
regions (where they are more difficult to observe; Minniti 1995). 
Elliptical galaxies reveal a
range of ratios, but the metal--poor almost always exceed the number of 
metal--rich ones (e.g. Forbes, Brodie \& Grillmair 1997). 

Mass is generally conserved in a merger, and
so from the total mass of the progenitor galaxies we can estimate the mass of
the resulting elliptical galaxy. 
By adding together the luminosities of all the galaxies in the Local Group
we derive a total magnitude of M$_V$ = --22.0. Thus if all galaxies in the
Local Group suddenly merged today, without any change in their overall
luminosity, we would end up with a highly luminous elliptical of 
M$_V$ = --22.0. 

Of course, not all of the stars in the LG galaxies resemble the old stellar
populations found in ellipticals, but they include intermediate and
young stellar populations. To generalise, the bulges of the spirals, 
dwarf ellipticals and spheriodals contain only old stars, whereas the 
disks of spirals and irregular galaxies contain intermediate and
young stellar populations. The detailed star formation histories of LG
galaxies are summarised by Grebel (1997). 

To crudely correct for this effect, we have  
separated the LG stars into `young' and `old' sub--populations.
For the three spirals we use 
bulge-to-disk luminosity ratios of 0.33, 0.24 and 0.09 for Sb (M31),
Sbc (Milky Way) and Sc (M33) respectively (Simien \& de Vaucouleurs 1986). 
The spiral bulges, dE, Sph and dSph galaxies (see Table 1) are 
included in the `old' sub--population and all the Irr and dIrr galaxies
as `young'. This gives the luminosity of the `old' LG stars as 
1.1 $\times$ 10$^{10}$ L$_{\odot}$  and the luminosity of the `young' stars
to be 4.2 $\times$ 10$^{10}$ L$_{\odot}$. Assuming a mass-to-light ratio
(M/L$_V$) of 5 for the old sub--population and 1 for the young
one  
gives masses of 5.7 $\times$ 10$^{10}$ M$_{\odot}$ 
and 4.1 $\times$ 10$^{10}$ M$_{\odot}$ respectively. This suggests that of
the stellar mass in the LG is divided fairly evenly between 
very old and more recently formed stars. 
In order to `simulate' the LG elliptical we convert the mass
of the `young' sub--population into a V band luminosity using 
the old stars mass-to-light ratio (i.e. M/L$_V$ = 5), thus giving it the 
luminosity that it would contribute to the final elliptical galaxy after
the stars had become comparably old. Combining this luminosity
with that of the LG old stars gives a total of M$_V$ = --20.9. This
luminosity is therefore more relevant for comparison to that of current day
ellipticals. In fact, the luminosity will be somewhat brighter than M$_V$ =
--20.9, as we have included the intermediate aged sub--population as `young'.

In Table 5 we list the specific frequency of the LG Elliptical
with those for other spiral galaxies. 
The LG Elliptical has 
N$_{GC}$ = 692 $\pm$ 125, corresponding to 
S$_N$ = 1.1 $\pm$ 0.2 for the total LG luminosity 
and S$_N$ = 3.0 $\pm$ 0.5 for the LG Elliptical 
using the stellar population corrected 
luminosity. The former value is comparable to those of the other spiral
galaxies in Table 5. The latter value is similar to the lower bound
of the range of S$_N$ for ellipticals, i.e. 3--6 (Harris 1991). This
suggests it is 
possible that the ellipticals at the lower bound of S$_N$ are 
the result of 
dissipationless mergers of galaxies containing 
young/intermediate stellar populations. 

The total stellar mass of the LG galaxies is 9.8 $\times$ 10$^{10}$
M$_{\odot}$ as calculated
above. Comparison with the total (dynamical) mass of the Local Group of 
2.3 $\pm$ 0.6 $\times$ 10$^{12}$ M$_{\odot}$ (Courteau \& 
van den Bergh 1999) indicates that only
about 5$\%$ of the LG mass is in stars. 

\section{Spectroscopic Metallicities for Globular Clusters in 
Elliptical Galaxies}
\label{Spectroscopic Metallicities for Globular Clusters in 
Elliptical Galaxies}

It is interesting to compare the results of section 4 with
ellipticals that have spectroscopically observed GCs. 
Only two early type galaxies outside of the LG have published
{\it individual}  
[Fe/H] determinations from spectroscopy. They are NGC 1399             
(Kissler--Patig et al. 1998) and M87 (Cohen et al. 1998), which are both cD
galaxies with a high specific frequency. In both cases the 
spectra were taken with the Keck 10m telescope. The GC metallicity 
distributions for these galaxies are shown in Fig. 4. The number of GCs
studied is only a small fraction of the total population, but assuming they
are representative, it appears that overall their GCs are slightly 
more metal--rich
on average than the LG elliptical. 

We note that in the case of M87, 
the GC system is clearly bimodal in its optical colours (e.g. Kundu et al.
1999). For the spectroscopic sub--sample of Cohen et al. (1998), the
distribution is not obviously bimodal to the eye as displayed in Fig. 4.  
If we exclude the 14 GCs with extreme metallicities, as done by Cohen 
et al. (1998), a KMM analysis indicates that a 
bimodal distribution (with peaks at [Fe/H] = --1.3 and --0.7) is 
preferred over a unimodal one at the 89$\%$
confidence level. If we use the colours of the two peaks from Kundu et al.
(1999), i.e. $V-I$ = 0.95 and 1.20, and the improved transformation from
Kissler--Patig et al. (1998), we derive values of [Fe/H] = --1.39 and
--0.59, which is reasonably consistent with the values from the
spectroscopic sample of Cohen et al.

\section{Concluding Remarks}
\label{Concluding Remarks}

We have collected together individual metallicity values for
globular clusters in the Local Group. Only the three large spirals show the
presence of a significant metal--rich (i.e. [Fe/H] $\sim$ --0.5) population of
globular clusters. Most galaxies have a metal--poor population with 
[Fe/H] $\sim$ --1.5. 
The GC systems of the Milky
Way and M31 appear to be typical examples of GC systems around spirals. 

The merger of the Local Group galaxies is expected to form an elliptical
galaxy. After the young stellar populations have faded, the resulting
galaxy will have M$_V$ $\le$ --21.0. 
If there is no creation or destruction of globular clusters, this
Local Group `Elliptical' will have about 700 globular clusters. The GC
system 
luminosity function will resemble the `universal' one. The metallicity
distribution will have peaks at [Fe/H] $\sim$ --1.55 and --0.64 with the
metal--poor to metal--rich ratio of 2.5:1. These peaks 
have a similar value
to that for the two early
type galaxies (M87 and NGC 1399) with available 
spectroscopically--determined 
metallicity distributions. The relative population ratio is also similar.
After a crude correction for stellar
populations, the specific frequency for the Local Group Elliptical is 
about 3. This value is similar to that for field and loose 
group ellipticals. 

From our `dissipationless merger simulation' we speculate that the globular
cluster systems
around low specific frequency 
ellipticals may be consistent with the simple combination
of the progenitor galaxies' globular cluster systems, 
and with little or no globular cluster formation. However the ellipticals
with moderate to high specific frequencies (i.e. S$_N$ = 5--15) require 
a higher globular cluster to field star ratio. In principle, this could be
solved with a dissipational merger (i.e. the creation of new globular clusters
from gas as seen in currently merging spirals). However the 
metal--rich globular cluster 
populations in these ellipticals are not larger in number than the 
metal--poor ones, as might be expected for the merger 
scenario (Forbes, Brodie \& Grillmair 1997). 
Alternatively, ellipticals and spirals alike, may form the vast bulk of
their metal--rich and
metal--poor globular clusters during two similar 
{\it in situ} formation episodes, as part of a dissipational 
galaxy collapse process. 

\acknowledgements{
We thank J. Brodie, G. Hau, J. Huchra 
for comments and useful discussion. We also thank the referee 
(J. Dubinski) for his comments. 
DM is supported by Fondecyt grant No. 01990440 and DIPUC, and by
the U.S. Department of Energy through Contract W-7405-Eng-48 to the
Lawrence Livermore National Laboratory. 
PB was supported by the Smithsonian Institution. 
}

\newpage

\begin{figure} 
\centerline{\psfig{figure=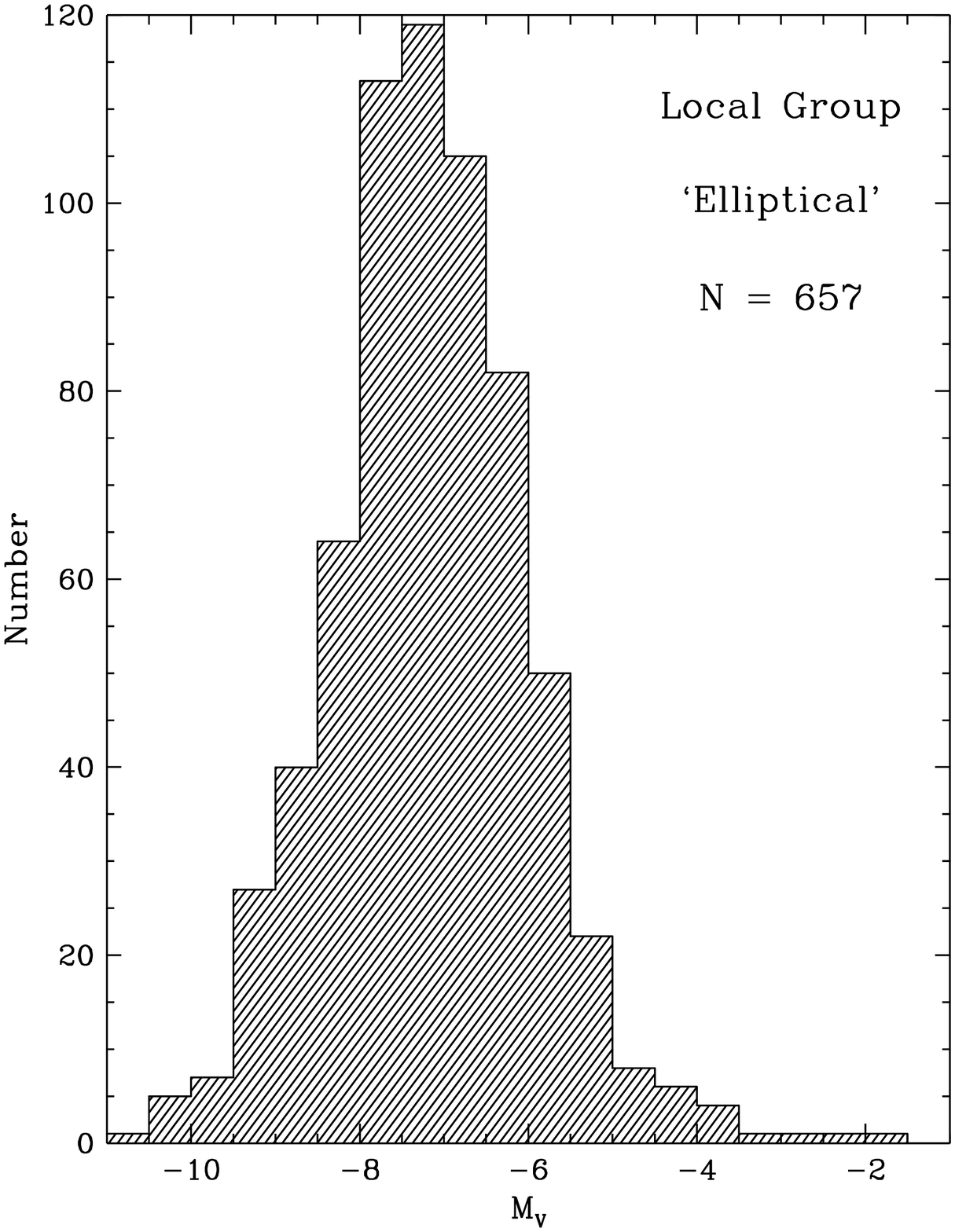,width=6in,height=6in}}
\caption{\label{fig1} Absolute magnitude distribution for the 
Local Group `Elliptical'. The distribution resembles the `universal'
globular cluster luminosity function. 
}
\end{figure}

%\begin{figure} 
%\centerline{\psfig{figure=mag.ps,width=6in,height=6in}}
%\caption{\label{fig1} Absolute magnitudes of Local Group globular clusters.
%The number of available globular clusters with individual measurements is 
%indicated in each panel.  
%}
%\end{figure}

\begin{figure} 
\centerline{\psfig{figure=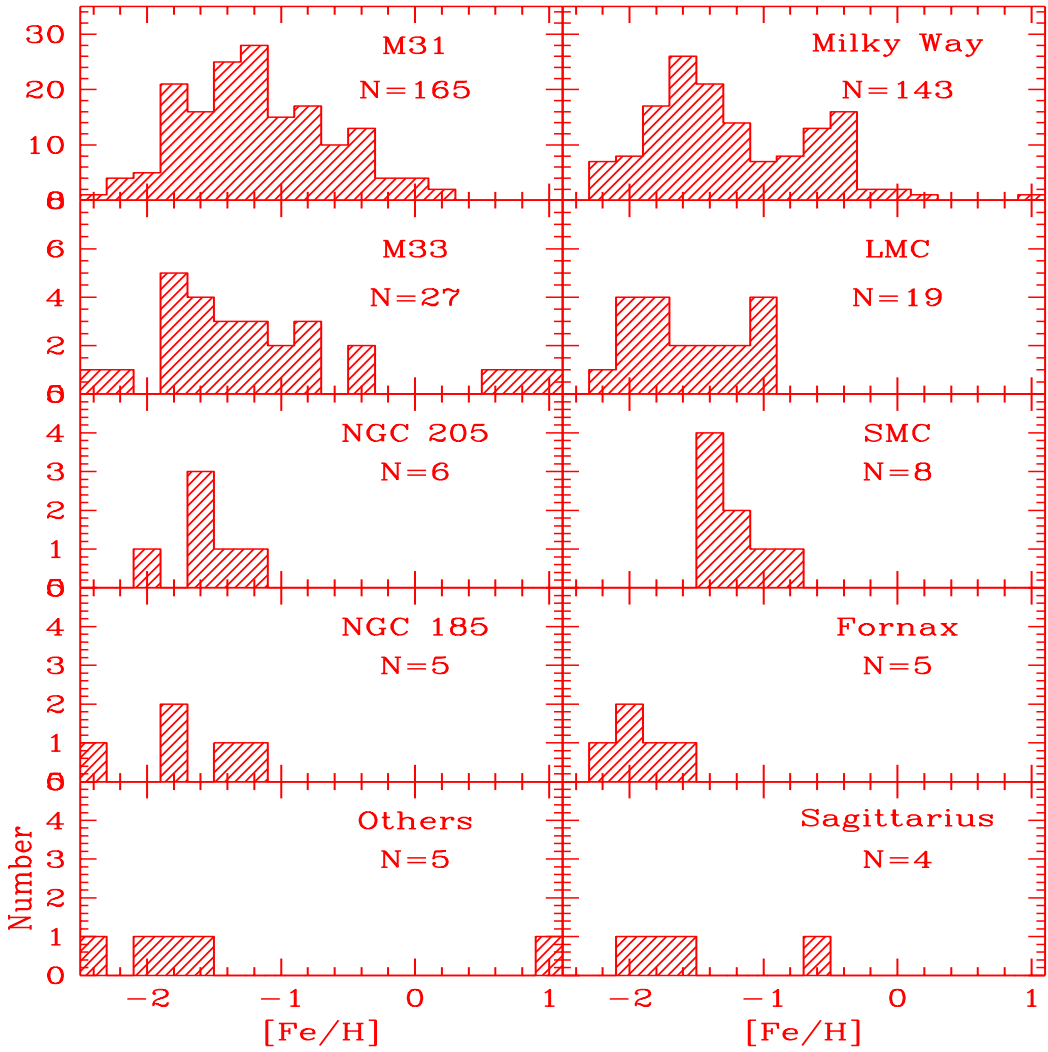,width=6in,height=6in}}
\caption{\label{fig2} Metallicities of Local Group globular clusters.
The number of available globular clusters with individual measurements is 
indicated in each panel.  
}
\end{figure}

%\begin{figure} 
%\centerline{\psfig{figure=n4565.ps,width=6in,height=6in}}
%\caption{\label{fig3} Metallicity distribution for globular clusters in the 
%Sb spiral NGC 4565. The B--I colours from Kissler--Patig et al. (1999) have 
%been converted into [Fe/H] based on a Galactic transformation. 
%}
%\end{figure}

\begin{figure} 
\centerline{\psfig{figure=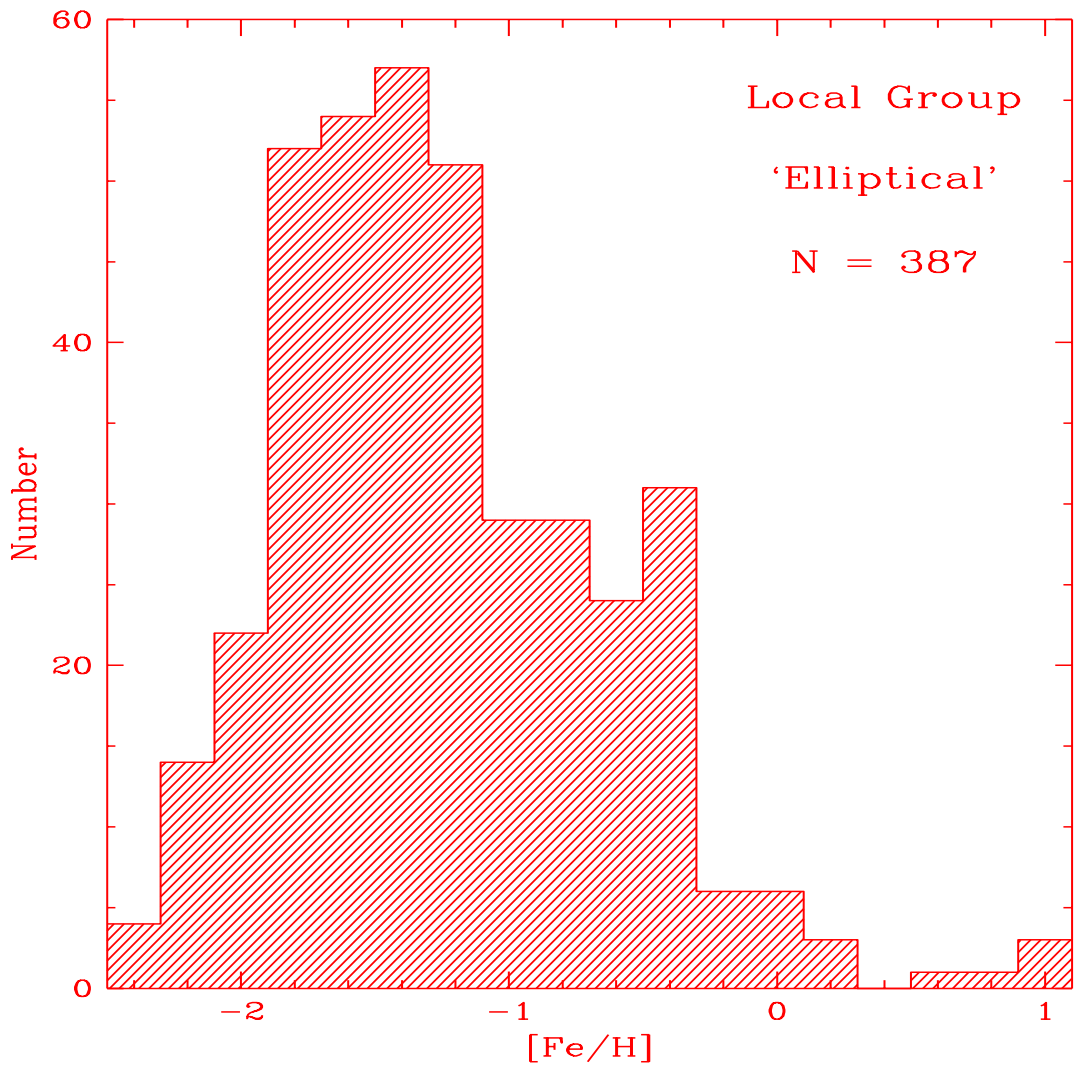,width=6in,height=6in}}
\caption{\label{fig3} Metallicity distribution for the 
Local Group `elliptical'. The distribution reveals two peaks at 
[Fe/H] = --1.55 and --0.64. 
}
\end{figure}

\begin{figure} 
\centerline{\psfig{figure=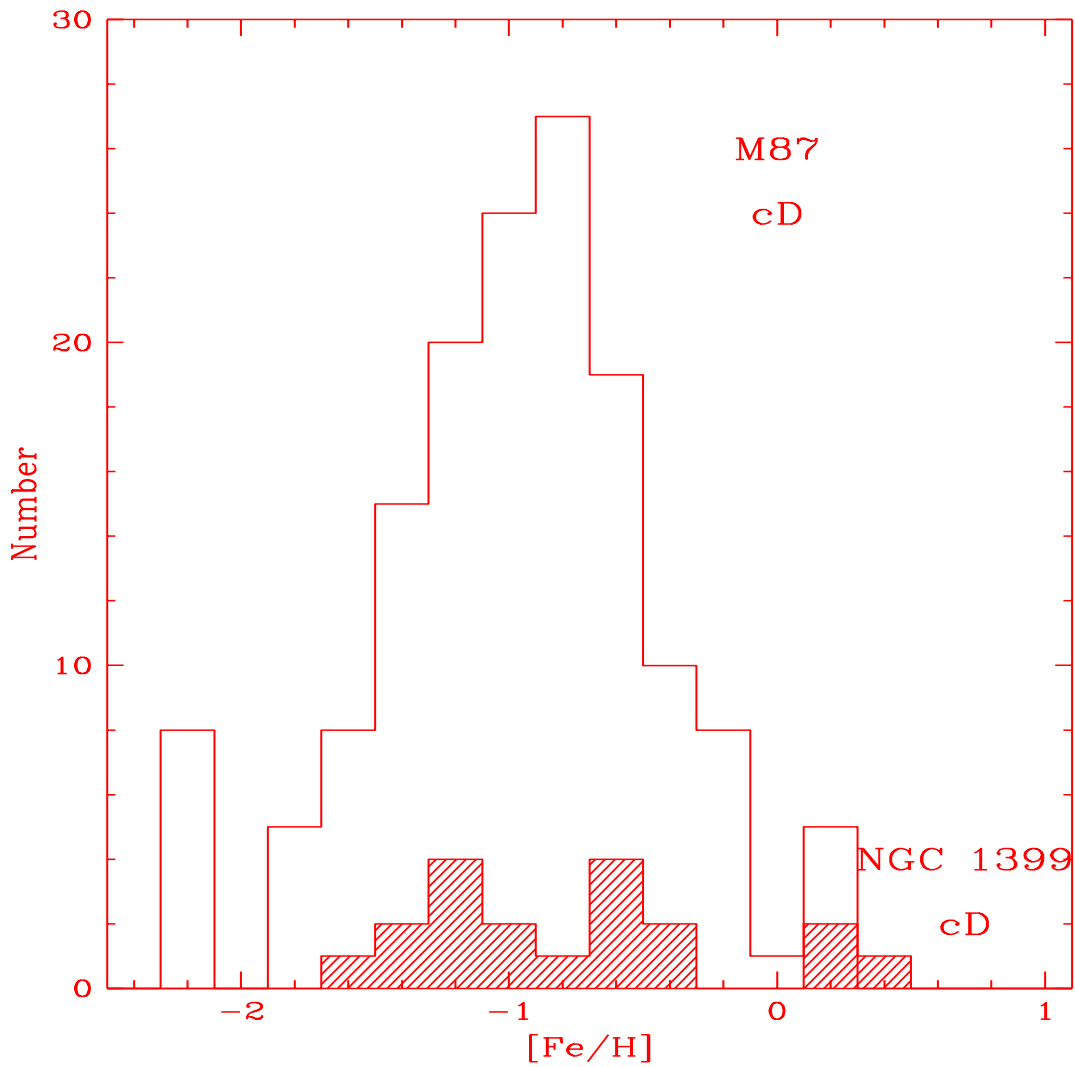,width=6in,height=6in}}
\caption{\label{fig4} Metallicity distribution for globular clusters in the 
cD galaxies M87 and NGC 1399 from Keck spectroscopy (Cohen et al. 1998; 
Kissler--Patig et al. 1998). 
}

\end{figure}

\end{document}